\documentclass[onecolumn]{aa}
\usepackage{epsfig,ifthen}

\def\vr{{\bf r}}
\def\mg{{\big <}}
\def\md{{\big >}}

\def\d{{\bf d}}

\def\zb{\overline{z}}
\def\vgammab{\overline{\gamma}}
\def\vgamma{{\bf \gamma}}
\newcommand{\Mfunction}[1]{{\ifthenelse{\equal{Gamma}{#1}}{\Gamma}{#1}}}
\def\Muserfunction#1(#2,#3){
{\ifthenelse{\equal{R1}{#1}}{G^{(#2)}(#3_3)}{}}
{\ifthenelse{\equal{R1b}{#1}}{{G}^{(#2)}(\zb_3)}{}}
{\ifthenelse{\equal{R2}{#1}}{H^{(#2)}(#3_3)}{}}
{\ifthenelse{\equal{R2b}{#1}}{{H}^{(#2)}(\zb_3)}{}}
}
\def\Mvariable#1{{\ifthenelse{\equal{zb}{#1}}{\zb_3}{}}}
\def\Ca{{C_{\alpha}(l_3)}}

\begin{document}

\title{Explicit computation of shear three-point correlation functions: the one-halo model case}
\titlerunning{the shear three-point correlation function in the one-halo model}
\author{ Francis Bernardeau}
\institute{ Service de Physique Th{\'e}orique,
             CEA/DSM/SPhT, Unit{\'e} de recherche associ{\'e}e au CNRS,
             CEA/Saclay F-91191 Gif-sur-Yvette cedex, France.}

\date{Received  / Accepted }

\abstract{
We present a method for calculating explicit expressions of the shear three-point function for various cosmological models. The method is applied here to the one-halo model in case of power law density profiles for which results are detailed. The three-point functions are found to reproduce to a large extent patterns in the shear correlations obtained in numerical simulations and may serve as a guideline to implement optimized methods for detecting the shear three-point function. 
In principle, the general method presented here can also be applied for other models of matter correlation.
\keywords{Cosmology: theory, dark matter, large-scale structure of Universe }
}
\maketitle

\section{Introduction}
Cosmic shear surveys offer a unique and unbiased view on the way matter cluster at cosmological scales and can then serve as a mean to explore the details of the gravitational instabilities. In particular the statistical properties of the shear field, in particular their high order correlation functions, are expected to reflect those of the matter density field (see Mellier 1999). The measurement of correlation functions in the cosmic shear surveys has then been recognized as a mean to break the
parameter degeneracy between the density parameter of the Universe and $\sigma_8$ that measurements of the shear power spectra
provide for (Bernardeau et al. 1997, Jain \& Seljak 1997) or a way to test the gravity law at cosmological scales (Bernardeau 2004). 

Matter correlation functions have been explored extensively in different regime. A reasonable understanding of their behaviors have been obtained in the quasilinear regime from perturbation theory techniques and into the nonlinear regime from phenomenological approaches~(see review of Bernardeau et al. 2002).
Recently a semianalytic technique based  on the halo model (Seljak 2000, Ma \& Fry 2000, Scoccimarro et al. 2001, see Cooray \& Sheth 2002 for a recent review) has attracted a lot of attention because it can provide accurate predictions for the galaxy or matter correlation properties.  So far however most of these theoretical approaches have dealt with the matter density field or with the convergence field. The later is closely related to the density field, being a simple projection of it, but the actual observational situation is more cumbersome. What is directly measured is the shear field (through galaxy shape measurements, see Mellier 1999 for a detailed review on the observational techniques) and it reveals difficult to build reliable projected density maps. 
It forces us to explore more deeply the high-order {\sl shear} correlation patterns. Those, because of their geometrical nature, are more intricate than that of the convergence maps. In particular complex relations are expected between the correlation components because both of
pseudo-vector nature of those and because one expects vanishing
"B" modes in the shear field. As a results shear correlation
coefficients depend in a complicated way on the basis on which
they are computed, have frequent sign changes, etc. (see Schneider and Lombardi 2003, for an exhaustive exploration of these properties). It makes the
construction of methods for measuring the shear three-point
function difficult. Attempts however have been made to detect the shear or aperture mass high order correlation functions (Bernardeau et al. 2003, Pen et al. 2003, Jarvis et al. 2004). Although these methods can give robust results, they may be far from optimal and therefore not completely satisfactory. One aim of this paper is to provide better foundations for the design of new methods for measuring shear three-point functions. 

In this paper the focus will be put on the one-halo model case for the description of the matter correlation properties. The idea is not however to build an accurate model for the mass distribution but to build a model good enough to reproduce the generic geometrical properties of the shear correlation patterns. The 1-halo model, with a power law density profile, is a good candidate for such a task being simple enough to allow complete computations of the correlation functions. 
The paper is divided as follows. The second section is devoted to a presentation of the mathematical tools applied to the one-halo model in the third section. The fourth section gives some insights on the results that have been obtained.

\section{A complex variable representation}

What we are interested in are the statistical properties of the either the projected density field $\phi(\vr)$, the convergence field $\kappa(\vr)$ and the 2-component shear field $(\gamma_{1},\gamma_{2})$. In the following the plane wave approximation is used, as usual, so that the convergence and shear field are all obtained through derivative operators applied to the projected potential,
\begin{eqnarray}
\kappa(\vr)&=&(\partial_{a}^{2}+\partial_{b}^{2})\,\phi(\vr)\\
(\gamma_{1},\gamma_{2})&=&\left(\partial_{a}^{2}-\partial_{b}^{2},2\partial_{a}\partial_{b}\right)\,\phi(\vr)
\end{eqnarray} 
where $\vr$ is the angular position of the point on the sky and $a$ and $b$ are the point coordinates.

In principle the convergence or shear correlations are functions of the point coordinates. Given the statistical isotropy and homogeneity of the fields those correlation functions should depend only on the relative positions and are globally rotationally invariant. Those correlation functions can therefore be written for instance as a function of the three distances of equivalently as a function of 2 distances and one angle, etc. 
The coordinate system advocated in this paper, and the one I will exploit throughout, is however none of those. Let me denote the point coordinates as $\vr_{1}$, $\vr_{2}$ and $\vr_{3}$ respectively ; $l_{1}$, $l_{2}$ and $l_{3}$ are the three lengths of the triangle they form such that $l_{3}^{2}=\vert \vr_{2}-\vr_{1}\vert^{2}$, etc. Then I can define the two ratios,
\begin{equation}
x_3=\frac{l_1^2}{l_3^2};\ y_3=\frac{l_2^2}{l_3^2}
\end{equation}
and the complex variable
\begin{equation}
z_3= \frac{1}{2}\left(1-x_3+y_3-\sqrt{(1-x_3+y_3)^2-4y_3}\right).
\end{equation}
It should be noted that $(1-x_3+y_3)^2-4y_3\le 0$ so that $z_3$ has generically a non vanishing imaginary part. The idea is then to write all correlation functions as a function of $(l_{i}, z_{i})$, $i$ being either 1, 2 or 3.  It is first to be remarked that $z_i$ is nothing but the components of $\vr_i$ in the basis defined by the other two points
\begin{equation}
z_3\ \to\ (a_3-{\rm i}\, b_3)/l_{3}.
\end{equation}
As a result the derivatives with respect to point coordinates can simply be re-expressed
as derivatives with respect to $z_{i}$ and $\zb_{i}$:
\begin{equation}
\partial_{\vr_{i}}=\frac{1}{l_{i}}(\partial_{z_{i}}+\partial_{\zb_{i}},-{\rm i}\,\partial_{z_{i}}+{\rm i}\,\partial_{\zb_{i}})
\end{equation}
It implies that the Laplacian operator, or operators that give access to the shear coordinates out of the potential field, respectively read
\begin{eqnarray}
\partial_{a}^{2}+\partial_{b}^{2}&=&\frac{4}{l^2}\partial_{z}\partial_{\zb}\\
\partial_a^2-\partial_b^2&=&\frac{2}{l^2}\left(\partial_{z}^2+\partial_{\zb}^2\right)\\
2\partial_a\partial_b&=&-\frac{2{\rm i}}{l^2}\left(\partial_{z}^2-\partial_{\zb}^2\right).
\end{eqnarray}
Clearly those equations suggest that the shear field is better written as a complex number 
$\vgamma=\gamma_{1}+{\rm  i}\gamma_{2}$ and we then have
\begin{eqnarray}
\kappa(l_{i},z_{i})&=&\frac{1}{l^2}\partial_{z_{i}}\partial_{\zb_{i}}\phi(l_{i},z_{i})\\
\vgamma(l_{i},z_{i})&=&\frac{1}{l^2}\partial_{z_{i}^2}\phi(l_{i},z_{i})\ \ {\rm and}\ \ {\vgammab}(l_{i},z_{i})=\frac{1}{l^2}\partial_{\zb_{i}^2}\phi(l_{i},z_{i})
\end{eqnarray}

Full computations with such variables will however been tractable only if change coordinates can be written in a natural way. Indeed the transform rules that give the $z_1$ and $z_2$,
complex coordinates of $\vr_1$ and $\vr_2$, as a function of $z_3$ are actually simple to compute and read,
\begin{equation}
z_1=\frac{1}{1-z_3},\ \ z_2=\frac{1}{1-z_1}=1-\frac{1}{z_3}.
\end{equation}
The coordinate change rules have to be completed with the following relations,
\begin{equation}
l_1^2=l_3^2\, z_1 \zb_1,\ \ \ l_2^2=l_1^2\,z_2\zb_2
\end{equation}
that gives the transformation rules for the base lengths. These rules suffice for the computations of correlation of scalar quantities, e.g. correlations between the projected potential and the convergence fields. 

For the computation of correlations between shear components however, further transform rules are required. These are the ones that rotate the shear coordinates from one basis to another. After a bit of vector analysis, these changes of coordinates simply amount to multiply the complex shear coordinates by $(1-z_3)/(1-\zb_3)$ in order to change base
$(\vr_1,\vr_2)$ to $(\vr_2,\vr_3)$ or by similar operations for the other changes (applying circular permutations to the indexes).

If the projected potential correlation function is known, then the computation of either the convergence field or the shear field is straightforward. What it takes is only a series of proper derivatives and rotation operators applied to the potential correlation function. 

This approach is particularly attractive when the latter correlation function can be written as a sum of terms of the form $f(z)\,f(\overline{z})$. In this case indeed the operators act independently on each factor. As we will see next this is precisely the situation that is encountered in the one-halo model with power law profiles.

\section{Correlation functions in the 1-Halo model} 

The halo model provides with a simplified description of the matter correlation properties. It has been advocated to be an accurate approach for describing the galaxy and matter correlation properties and studies of the dark matter density field through halo contributions have recently been developed in Seljak (2000), Ma \& Fry (2000), Scoccimarro et al. (2001),
but also applied to three-point lensing statistics (Cooray et al. 2000b and Takada 
\& Jain 2003).

In this section we will illustrate the methods sketched in the previous section for shear three-point functions using the halo model restricting for simplicity to the one-halo term. The latter is indeed expected to dominate on the small scales (see for instance Fosalba et al. 2005) where the three-point function of the shear is easiest to measure in observations. In itself the use of the halo model for the computation of the lensing high order correlations is not new. It has been done by Cooray \& Hu (2001) for the convergence bispectrum and in Takada \& Jain 2002a, Zaldarriaga \& Scoccimarro (2003) for the shear correlation functions. The accuracy of the model and the has been extensively commented in those papers and in Fosalba et al . (2005). It won't be commented any further it here.

The basic relation for computing the correlation functions in the one-halo model has been provided for by Scoccimarro (1997, taking advantage of a result given by Davydychev 1992) as discussed in Zaldarriaga and Scoccimarro (2003).

The one-halo amounts to assume that a stochastic field $\phi$ can be represented by a randomly located halo with a spherically symmetric $\phi$ profile, $\phi(\vr)\sim F(\vert\vr-\vr_0\vert)$ where $\vr_0$, the center of the halo is evenly distributed in space. The $\phi$ correlation functions then read,
\begin{equation}
F_n(\vr_1\dots\vr_n)=\mg \phi(\vr_1)\dots\phi(\vr_n)\md \sim
\int\d^d\vr_0\, F(\vert\vr_1-\vr_0\vert)\dots
F(\vert\vr_n-\vr_0\vert)
\end{equation}

This expression is in general complicated. We are interested in such an expression for the three-point function and in case of a
power law profile, $F(x)\sim x^{-\alpha}$. It is to be noted that such an integral converge for $d-3\alpha<0$ and $d-\alpha>0$. For
$F(x)=F_0\,x^{-\alpha}$ the final result can be given in a closed form in terms of hypergeometric functions. This result has been
presented by Scoccimarro (1997) for the three-point function taken advantage of integrals developed in the context of high energy physics.

We are interested in this result in the context of a 2D projected density field\footnote{The scales at which the correlations are to
be measured are small enough so that the small angle approximation can be used.}. The result reads
\begin{eqnarray}
F_3(\vr_1,\vr_2,\vr_3)&=&l_3^{2-3\alpha}\frac{\pi^3}{2\sin(\alpha
\pi/2)^2\,\cos(\alpha
\pi/2)\,\Gamma(2-3\alpha/2)\Gamma(\alpha/2)^2}\times\\
&&\left[\Gamma(\alpha/2)\Gamma(1-3\alpha/2)G(z_3)\,{G}(\zb_3)
-\Gamma(2-3\alpha/2)\Gamma(1-\alpha/2)H(z_3){H}(\zb_3)\right]\label{F3}
\end{eqnarray}
where the variables are those defined in the previous part and where ${G}(z)$ and ${H}(z)$
are respectively,
\begin{eqnarray}
G(z)&=&{\left( 1 - z \right) }^{1 -
\alpha}\,\frac{\ _2\!F_1 \left(\frac{\alpha}{2},1 -
\frac{\alpha}{2},\alpha,z\right)}{\Gamma(\alpha)}
\\
H(z)&=&z^{1-\alpha}\,\frac{\
_2\!F_1\left(\frac{\alpha}{2},1 - \frac{\alpha}{2},2 -
\alpha,z\right)}{\Gamma(2-\alpha)}.
\end{eqnarray}

This expression results from a slight modification of the formula presented in Scoccimarro (1997). From the form (\ref{F3}) it is clear that $F_3$ is real.

The idea pursued in this paper is that convergence fields can be described to a reasonable accuracy with the 1-halo model with a
density profile close to $\alpha=1$. The goal though is not to use (\ref{F3}) for the convergence field but to construct in this
framework the shear three-point functions,
\begin{equation}
\xi_{\gamma_i\gamma_j\gamma_k}(\vr_1,\vr_2,\vr_3)=
\mg\gamma_i(\vr_1)\gamma_j(\vr_1)\gamma_k(\vr_1)\md
\end{equation}
where $i,j,k$ stand for the two component indices of each of the shear field.

To obtain this function two routes can be followed. The one advocated in Zaldarriaga and Scoccimarro (2003) was to transform the formal
expression of $\xi_{\gamma_i\gamma_j\gamma_k}$ into a sum of integral each of which can be computed in a manner similar to
$F_3$. It leads however to expressions with a very large number of terms that reveal hard to manipulate and simplify.

The route followed here is radically different. As suggested in the second section it amounts to take proper subsequent derivatives of the potential three-point
function, $\xi_{\phi\phi\phi}(\vr_1,\vr_2,\vr_3)$ with respect to the positions. The potential correlation function is a priori not
well defined because it requires the use of (\ref{F3}) when $\alpha$ is about -1. It demands then a regularization of the profile at
large distance. It can be checked however that the introduction of this cutoff does not affect the finite distance behavior of its
derivatives if those are finite.

The general mathematical operators to be used have been presented in the previous section. Least by not last, the calculations remain tractable in case of this model thanks to the fact that the function $G$ and $H$ have simple derivatives. It can indeed be shown
that (using Gradshteyn and Ryzhik 1980, Eqs. 9.137.6, 9137.7 and 9.137.17)
\begin{eqnarray}
\frac{d^n}{d z^n}G(z)&=&G^{(n)}(z)={\left( 1 - z \right) }^{1 -
\alpha-n}\,\frac{\ _2\!F_1 \left(\frac{\alpha}{2},1 -
\frac{\alpha}{2},\alpha+n,z\right)}{\Gamma(\alpha+n)}
\frac{\Gamma(\alpha/2+n)}{\Gamma(\alpha/2)}\frac{\Gamma(1-\alpha/2+n)}{\Gamma(1-\alpha/2)}\\
\frac{d^n}{d z^n}H(z)&=&H^{(n)}(z)=z^{1-\alpha-n}\,\frac{\
_2\!F_1\left(\frac{\alpha}{2},1 - \frac{\alpha}{2},2 -
\alpha-n,z\right)}{\Gamma(2-\alpha-n)}
\end{eqnarray}

To get the shear correlation functions the subsequent steps should be followed,
\begin{eqnarray}
\xi_{\phi\phi\gamma}(l_{3},z_{3})&=& \frac{1}{l_{3}^{2}}\partial_{z_{3}^{2}}\xi_{\phi\phi\phi}(l_{3},z_{3})\\
\xi_{\phi\gamma\gamma}(l_{2},z_{2})&=&\frac{1}{l_{2}^{2}}\partial_{z_{2}}^{2} \left\{R_{32} \xi_{\phi\phi\gamma}(l_{2},z_{2})\right\}\\
\xi_{\gamma\gamma\gamma}(l_{1},z_{1})&=&R_{13}^{3}\frac{1}{l_{2}^{2}}\partial_{z_{1}}^{2}\left\{ R_{21}^{2} \xi_{\phi\gamma\gamma}(l_{1},z_{1})\right\}
\end{eqnarray}
and similar expression for the other components of the shear correlation functions, $\xi_{\vgamma(\vr_1)\vgamma(\vr_2)\vgammab(\vr_3)}$, $\xi_{\vgamma(\vr_1)\vgammab(\vr_2)\vgammab(\vr_3)}$, etc.. Note that in those expressions a change of variable should be applied from one line to another (for both the $z$ and the $l$ variables) ; the rotation operators are  $R_{21}=1/R_{12}=(1-\zb_{1})/(1-z_{1})$, $R_{13}=1/R_{31}=(1-\zb_{3})/(1-z_{3})$, $R_{32}=R_{31}R_{12}=(1-z_{3})/(1-\zb_{3})\,(1-z_{1})/(1-\zb_{1})$ and they should be replaced by their inverse (or equivalently their complex conjugate) each time it applies to conjugate components of the shear.

After calculations and simplification taking advantage of recursion relations of the hypergeometric functions and with the help of formal calculators one can obtain explicit expressions for the shear correlation functions.  The first four read
\begin{eqnarray}
\xi_{\gamma(\vr_1)\gamma(\vr_2)\gamma(\vr_3)}&=&-\frac{\Ca\,{\alpha}^2\,{\left(
2 + \alpha \right) }^2}{{\left( -1 + z_3 \right) }^3\,z_3^3}\,
      \left\{ \left[ 2\,\alpha\,{\left( 1 - 2\,z_3 \right) }^2\,\left( -2 - z_3 + z_3^2 \right)  +
           8\,\left( 1 - 2\,z_3 + 4\,z_3^3 - 2\,z_3^4 \right)
           +\right.\right.\nonumber\\
&&\left.\left.        3\,{\alpha}^2\,
           \left( -4 + 5\,z_3 + 3\,z_3^2 - 16\,z_3^3 + 8\,z_3^4 \right)
           \right] \,\Mfunction{Gamma}(\frac{\alpha}{2})\,
           \Mfunction{Gamma}(-1 + \frac{3\,\alpha}{2})\,\Muserfunction{R1}(0,z)\,
         \Muserfunction{R1b}(0,z) +\right.\nonumber\\
&&\left.
          8\,\left( 1 + \alpha \right) \,\left( 2 - 5\,z_3 + 2\,z_3^2 + 2\,z_3^3 - 5\,z_3^4 + 2\,z_3^5 \right) \,
         \Mfunction{Gamma}(\frac{\alpha}{2})\,\Mfunction{Gamma}(-1 + \frac{3\,\alpha}{2})\,
         \Muserfunction{R1}(1,z)\,\Muserfunction{R1b}(0,z) -
\right.\nonumber\\
&&\left.         \Mfunction{Gamma}(2 -
\frac{3\,\alpha}{2})\,\Mfunction{Gamma}(1 - \frac{\alpha}{2})\,
         \left[ \left( 2\,\alpha\,{\left( 1 - 2\,z_3 \right) }^2\,\left( -2 - z_3 + z_3^2 \right)  +
              8\,\left( 1 - 2\,z_3 + 4\,z_3^3 - 2\,z_3^4 \right)  +
\right.\right.\right.\nonumber\\
&&\left.\left.\left.               3\,{\alpha}^2\,\left( -4 +
5\,z_3 + 3\,z_3^2 - 16\,z_3^3 + 8\,z_3^4 \right)  \right)
             \,\Muserfunction{R2}(0,z) +
\right.\right.\nonumber\\
&&\left.\left.
           8\,\left( 1 + \alpha \right) \,\left( 2 - 5\,z_3 + 2\,z_3^2 + 2\,z_3^3 - 5\,z_3^4 + 2\,z_3^5 \right)
           \,\Muserfunction{R2}(1,z)
           \right] \,
\Muserfunction{R2b}(0,z)  \right\}\\
\xi_{\gamma(\vr_1)\gamma(\vr_2)\overline{\gamma}(\vr_3)}&=& -
\frac{\Ca\,{\alpha}^2\,\left( 2 + \alpha \right)}{\left( -1 + z_3
\right) \,z_3\,
      \left( -1 + \Mvariable{zb} \right) \,\Mvariable{zb}} \,
      \left\{ -\left[ \Mfunction{Gamma}(\frac{\alpha}{2})\,\Mfunction{Gamma}(-1 + \frac{3\,\alpha}{2})\,
           \left( \left( -2 + 3\,\alpha \right) \,\left( 8\,\left( -1 + z_3 \right) \,z_3 +
\right.\right.\right.\right.\nonumber\\
&&\left.\left.\left.\left.                  \alpha\,\left( -1 - 8\,z_3 + 8\,z_3^2 \right)  \right)
\,\Muserfunction{R1}(0,z) +
             4\,\left( -1 + 2\,z_3 \right) \,\left( 2\,\left( -1 + z_3 \right) \,z_3 +
             \right.\right.\right.\right.\nonumber\\
&&\left.\left.\left.\left.\, \alpha\,\left( -1 - 2\,z_3 + 2\,z_3^2 \right)
                \right) \,\Muserfunction{R1}(1,z) \right)\,
           \left( \left( -2 + 3\,\alpha \right) \,\Muserfunction{R1b}(0,z) +
             4\,\left( -1 + 2\,\Mvariable{zb} \right) \,\Muserfunction{R1b}(1,z) \right)  \right]
             +\right.\nonumber\\
&&\left.        \Mfunction{Gamma}(2 -
\frac{3\,\alpha}{2})\,\Mfunction{Gamma}(1 - \frac{\alpha}{2})\,
         \left[ \left( -2 + 3\,\alpha \right) \,\left( 8\,\left( -1 + z_3 \right) \,z_3 +
              \alpha\,\left( -1 - 8\,z_3 + 8\,z_3^2 \right)  \right) \,\Muserfunction{R2}(0,z) +
\right.\right.\nonumber\\
&&\left.\left.           4\,\left( -1 + 2\,z_3 \right) \,\left(
2\,\left( -1 + z_3 \right) \,z_3 + \alpha\,\left( -1 - 2\,z_3 +
2\,z_3^2 \right)  \right)
              \,\Muserfunction{R2}(1,z) \right] \,\left( \left( -2 + 3\,\alpha \right) \,\Muserfunction{R2b}(0,z) +
\right.\right.\nonumber\\
&&\left.\left.
           4\,\left( -1 + 2\,\Mvariable{zb} \right) \,\Muserfunction{R2b}(1,z) \right)
           \right\}
      \\
\xi_{\gamma(\vr_1)\overline{\gamma}(\vr_2){\gamma}(\vr_3)}&=&
\frac{\Ca\,\alpha\,\left( 2 + \alpha \right)}{{\left( -1 + z_3
\right) }^3\,z_3\,\Mvariable{zb}} \,\left\{
-2\,\Mfunction{Gamma}(1 + \frac{\alpha}{2})\,
       \Mfunction{Gamma}(-1 + \frac{3\,\alpha}{2})\,
       \left[ \left( -2 + 3\,\alpha \right) \,\left( -4\,\Mfunction{z_3} + \alpha\,\left( -1 - 5\,z_3 + 2\,z_3^2 \right)  \right) \right.\right.\nonumber\\
&&\left.\left.  
\Muserfunction{R1}(0,z) +4\,\left( 1 + z_3 \right) \,\left( -2\,z_3 +
\alpha\,\left( 1 - 4\,z_3 + z_3^2 \right) \right) \,
          \Muserfunction{R1}(1,z) \right]\,
\times\right.\nonumber\\
&&\left.           
\left[ \left( -2 + 3\,\alpha \right) \,\left( 1 +
2\,\Mvariable{zb} \right) \,
          \Muserfunction{R1b}(0,z) + 4\,\left( -1 + {\Mvariable{zb}}^2 \right) \,\Muserfunction{R1b}(1,z) \right]
          +
\right.\nonumber\\
&&\left.
      \alpha\,\Mfunction{Gamma}(2 - \frac{3\,\alpha}{2})\,\Mfunction{Gamma}(1 - \frac{\alpha}{2})\,
       \left[ \left( -2 + 3\,\alpha \right) \,\left( -4\,\Mfunction{z_3} + \alpha\,\left( -1 - 5\,z_3 + 2\,z_3^2 \right)  \right) \,
          \Muserfunction{R2}(0,z) +
\right.\right.\nonumber\\
&&\left.\left.
          4\,\left( 1 + z_3 \right) \,\left( -2\,z_3 + \alpha\,\left( 1 - 4\,z_3 + z_3^2 \right)  \right) \,
          \Muserfunction{R2}(1,z) \right] \,
\times\right.\nonumber\\
&&\left.
          \left[ \left( -2 + 3\,\alpha \right) \,\left( 1 + 2\,\Mvariable{zb} \right) \,
          \Muserfunction{R2b}(0,z) + 4\,\left( -1 + {\Mvariable{zb}}^2 \right) \,\Muserfunction{R2b}(1,z)
          \right]
          \right\}\\
\xi_{\gamma(\vr_1)\overline{\gamma}(\vr_2)\overline{\gamma}(\vr_3)}&=&
\frac{\Ca\,\alpha\,\left( 2 + \alpha \right)}{\left( -1 + z_3
\right) \,\left( -1 + \Mvariable{zb} \right) \,{\Mvariable{zb}}^3}
\,\left\{ 2\,\Mfunction{Gamma}(1 + \frac{\alpha}{2})\,
       \Mfunction{Gamma}(-1 + \frac{3\,\alpha}{2})\,
       \left[ \left( -6 + \alpha\,\left( 9 - 6\,z_3 \right)  + 4\,z_3 \right) \,\Muserfunction{R1}(0,z) -
\right.\right.\nonumber\\
&&\left.\left.
         4\,\left( -2 + z_3 \right) \,z_3\,\Muserfunction{R1}(1,z) \right] \,
       \left[ \left( -2 + 3\,\alpha \right) \,\left( 4\,\left( -1 + \Mvariable{zb} \right)  +
            \alpha\,\left( -4 + \Mvariable{zb} + 2\,{\Mvariable{zb}}^2 \right)  \right) \,\Muserfunction{R1b}(0,z) +
\right.\right.\nonumber\\
&&\left.\left.
         4\,\left( -2 + \Mvariable{zb} \right) \,\left( 2\,\left( -1 + \Mvariable{zb} \right)  +
            \alpha\,\left( -2 + 2\,\Mvariable{zb} + {\Mvariable{zb}}^2 \right)  \right) \,\Muserfunction{R1b}(1,z) \right]  +
      \alpha\,\Mfunction{Gamma}(2 - \frac{3\,\alpha}{2})\,\Mfunction{Gamma}(1 -
      \frac{\alpha}{2})\,\times
\right.\nonumber\\
&&\left.      \left[ \left( -2 + 3\,\alpha \right) \,\left( -3 +
2\,z_3 \right) \,\Muserfunction{R2}(0,z) +
         4\,\left( -2 + z_3 \right) \,z_3\,\Muserfunction{R2}(1,z) \right]
         \times
\right.\nonumber\\
&&\left.
       \left[ \left( -2 + 3\,\alpha \right) \,\left( 4\,\left( -1 + \Mvariable{zb} \right)  +
            \alpha\,\left( -4 + \Mvariable{zb} + 2\,{\Mvariable{zb}}^2 \right)  \right) \,\Muserfunction{R2b}(0,z) +
\right.\right.\nonumber\\
&&\left.\left.
         4\,\left( -2 + \Mvariable{zb} \right) \,\left( 2\,\left( -1 + \Mvariable{zb} \right)  +
            \alpha\,\left( -2 + 2\,\Mvariable{zb} + {\Mvariable{zb}}^2 \right)  \right) \,\Muserfunction{R2b}(1,z)
            \right]
      \right\}
\end{eqnarray}
with
\begin{equation}
\Ca=\frac{-l_3^{-4 - 3\,\alpha}\left( {\pi }^3\,{\csc
(\frac{\alpha\,\pi }{2})}^2\,\sec (\frac{\alpha\,\pi }{2}) \right)
}
  {2\,\Mfunction{Gamma}(2 -
  \frac{3\,\alpha}{2})\,{\Mfunction{Gamma}(\frac{\alpha}{2})}^3}.
\end{equation}
The other four components can be derived as complex conjugates. The results are given as a function of $z_3$ and $\zb_3$ and
therefore when the coordinates (in particular the shear coordinates) are expressed in the $(\vr_1,\vr_2)$ basis.

\section{Shear Patterns in the One-Halo Model}

\begin{figure}{\vspace{.2in}
\centerline {
\includegraphics[width=6cm]{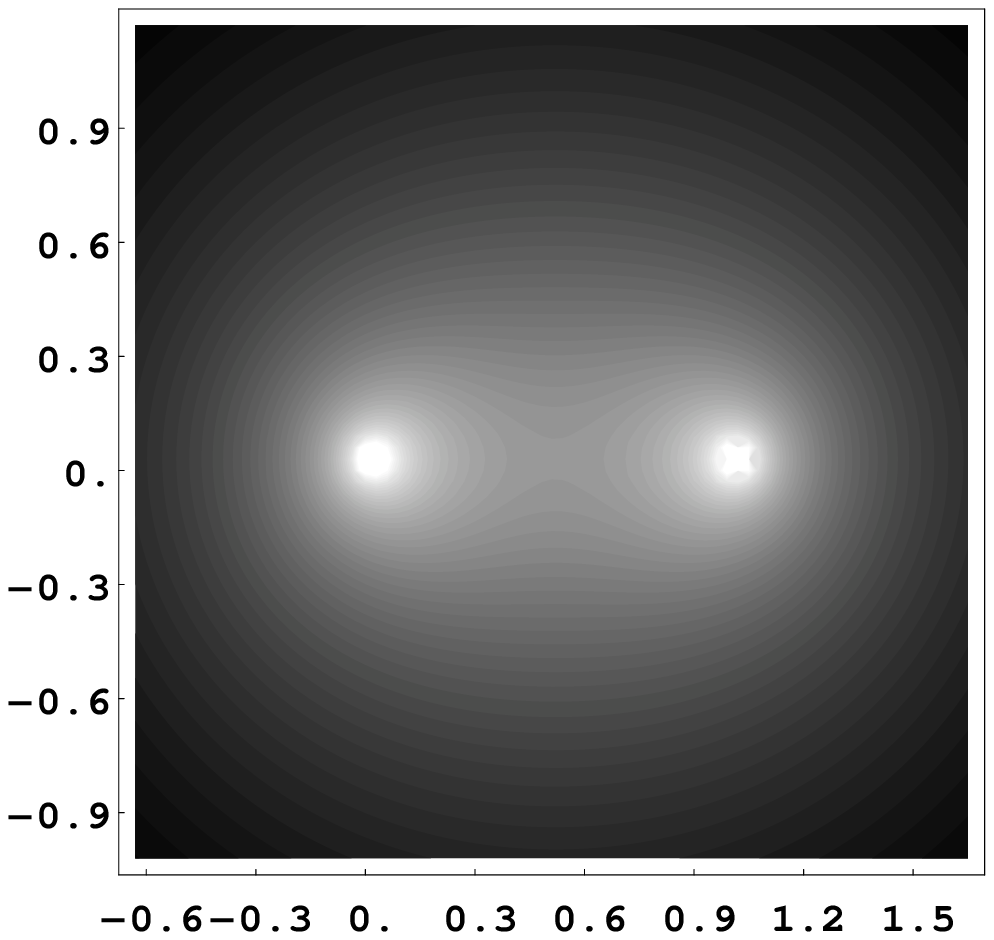}\hspace{1cm}
\includegraphics[width=6cm]{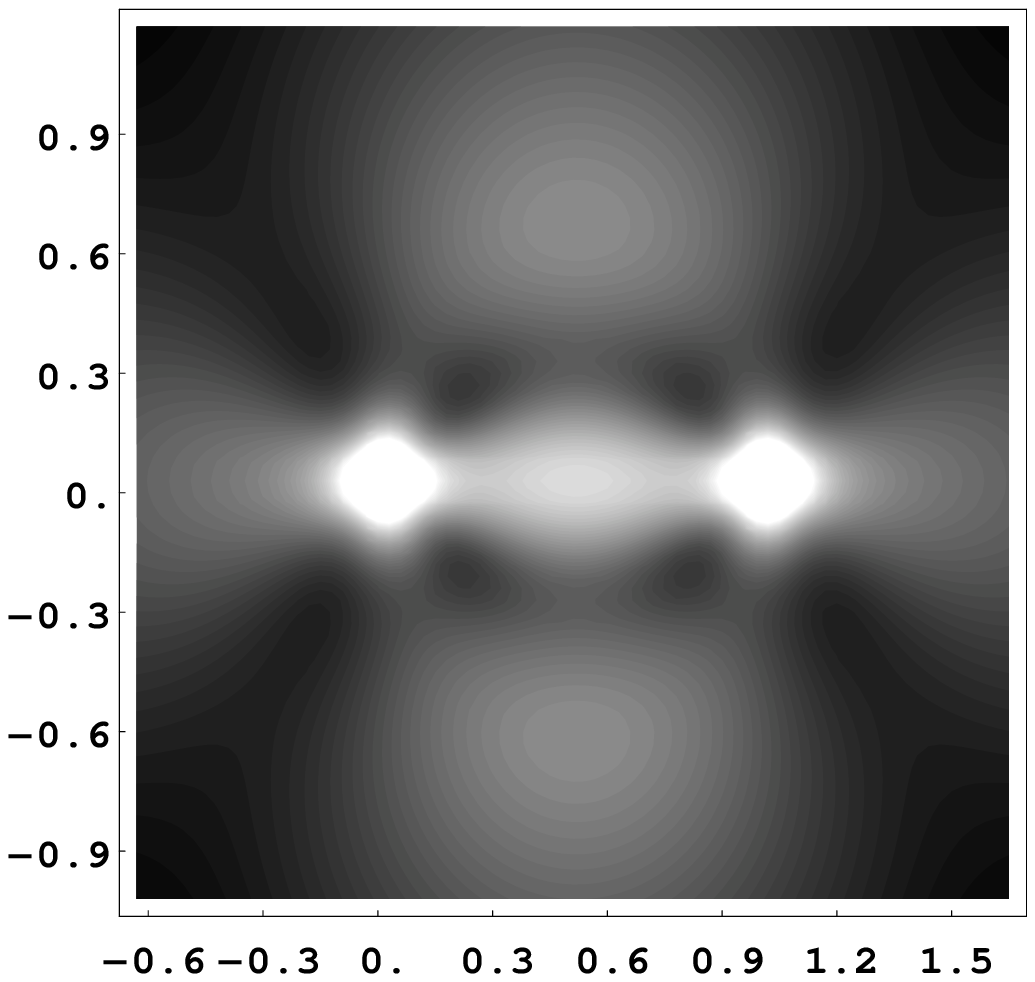}
}
\vspace{.2in}
}
\caption{Comparison of the spatial dependence of the amplitude of the three point function for both the convergence fied (left panel) and the shear field (right panel). These result correspond to the one-halo model with $\alpha=-1.$. In this plot, $\vr_{1}=(0,0)$ and $\vr_{2}=(1,0)$. As can be observed the spatial structure of the shear three-point function is much more complex than the one of the convergence field}
\label{xi3norm}\end{figure}

\begin{figure}{\vspace{.2in}
\centerline {
\includegraphics[width=6cm]{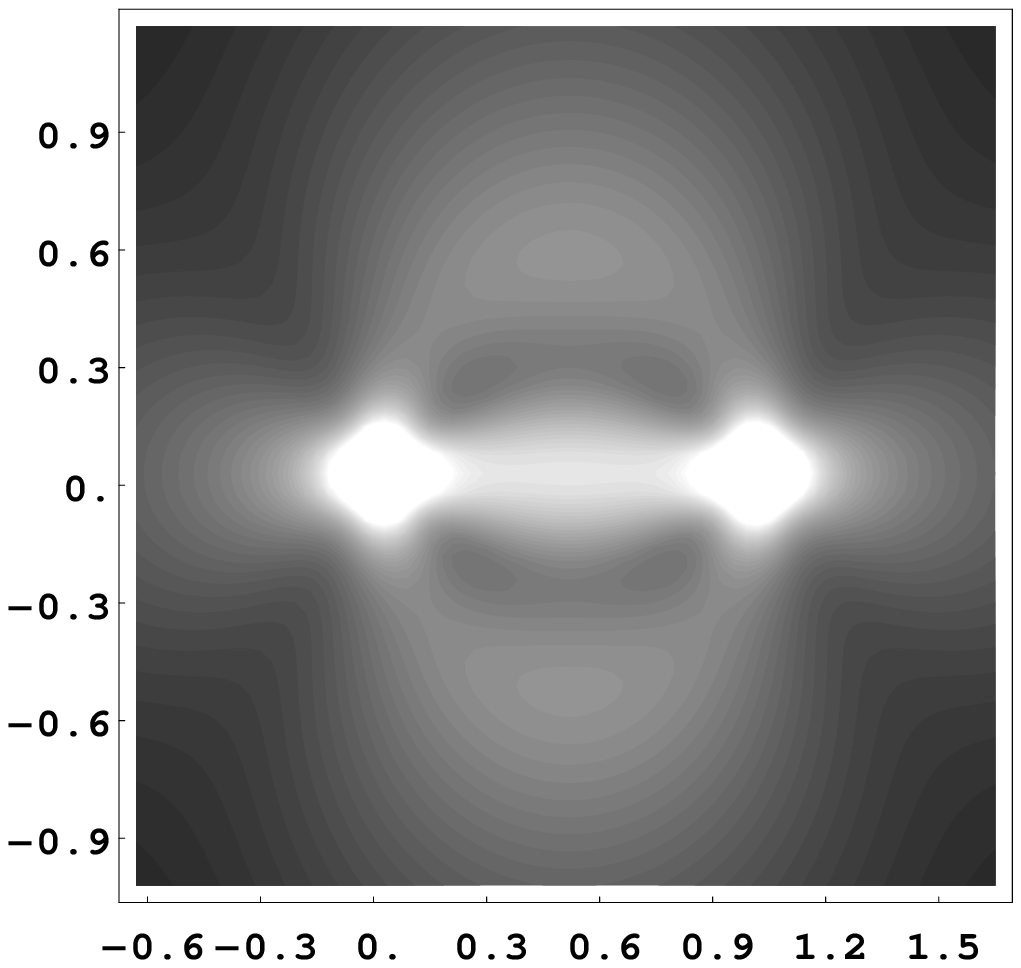}\hspace{1cm}
\includegraphics[width=6cm]{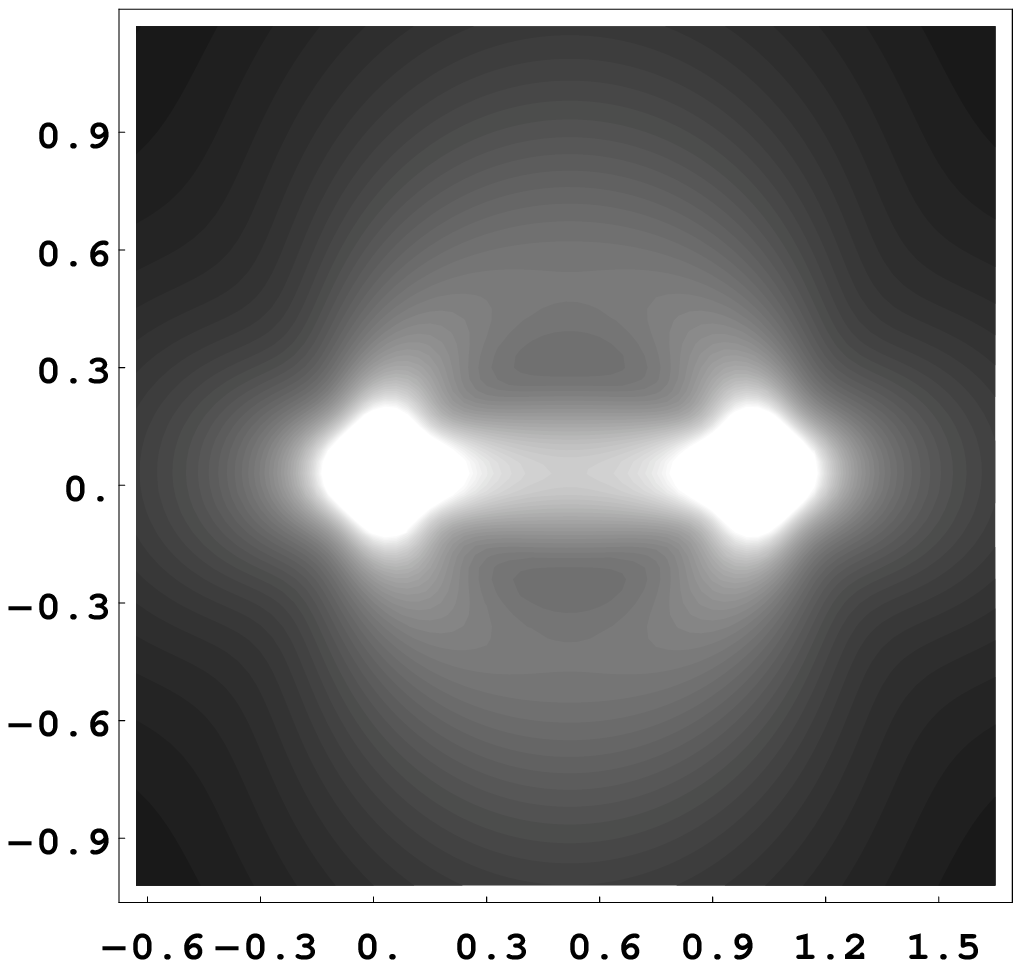}
}
\vspace{.2in}
}
\caption{Comparison of the spatial dependence of the amplitude of the three point function for $\alpha=-0.75$ (left panel) and $\alpha=-0.5$ (right panel). }
\label{xi3norm23}\end{figure}

The amplitude of the shear correlation function can then be compared to the amplitude of the convergence correlation function.
To start with one can consider $\xi_{\gamma}(z)$ defined as
\begin{equation}
\xi_{\gamma}(z)=\left(\sum_{ijk}\xi_{\gamma_{i}(\vr_{1})\gamma_{j}(\vr_{2})\gamma_{k}(\vr_{3})}\ \xi_{\gamma_{i}(\vr_{1})\gamma_{j}(\vr_{2})\gamma_{k}(\vr_{3})}\right)^{1/2}.
\end{equation}

It is to be noted that there are some symmetry properties in this quantity due to the fact it should be invariant under $z\to 1/(1-z)$ and $z\to 1-1/z$ when properly rescaled (these changes correspond to transform that leaves the triangle shape unchanged; the amplitude of the shear functions are then only changed by a scale factor effect).

Its amplitude in shown on Fig. \ref{xi3norm} where the spatial dependence of the amplitude of the shear correlation function is compared to that of the convergence correlation function. Whereas the convergence correlation function has a very simple spatial dependence, the spatial dependence of the shear correlation is much more complex exhibiting minima for certain configurations. 
It is to be noted however that these patterns are quite dependent of the power spectrum index (right panel of Fig. 1 and Fig. 2). In particular the existence of a local maximum of the signal for equilateral triangles vanishes for steep spectra.

A fair fraction of the signal however can be extracted from a subpart of the correlation functions. Following Bernardeau et al. (2003) it is possible to map at least some
aspects of the shear patterns while considering the function,
\begin{equation}
\xi_3(\vr_3)=\mg (\gamma_{\vr_1}\cdot\gamma_{\vr_2})\,\gamma_{\vr_3}\md
\end{equation}
which can be written in terms of the complex coordinate
correlations as
\begin{equation}
\xi_3(\vr_3)=\frac{1}{2}\left(
\xi_{\gamma(\vr_1)\overline{\gamma}(\vr_2)\gamma(\vr_3)}+
\xi_{\overline{\gamma}(\vr_1){\gamma}(\vr_2)\gamma(\vr_3)}\right)
\end{equation}

\begin{figure}{\vspace{.2in}
\centerline {
\includegraphics[width=6cm]{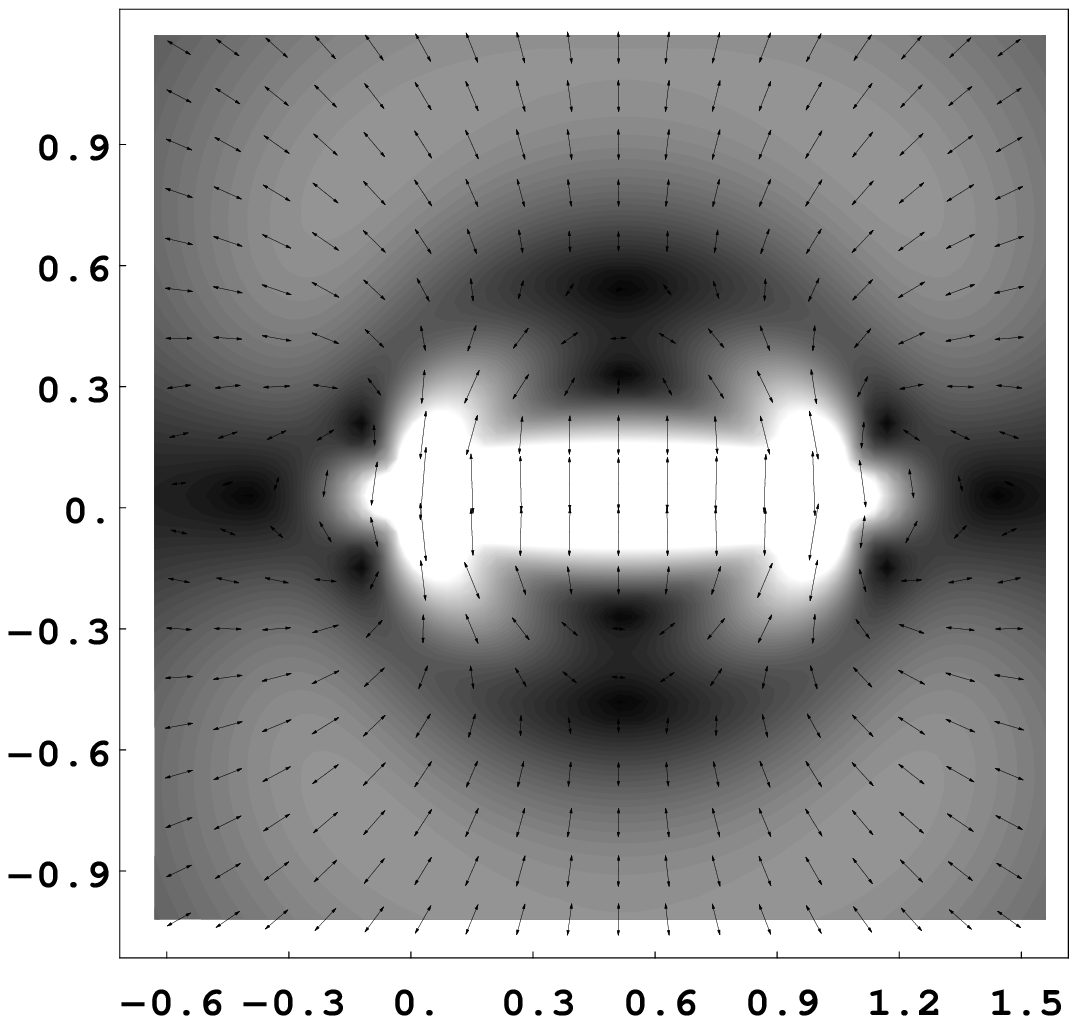}\hspace{1cm}
\includegraphics[width=6cm]{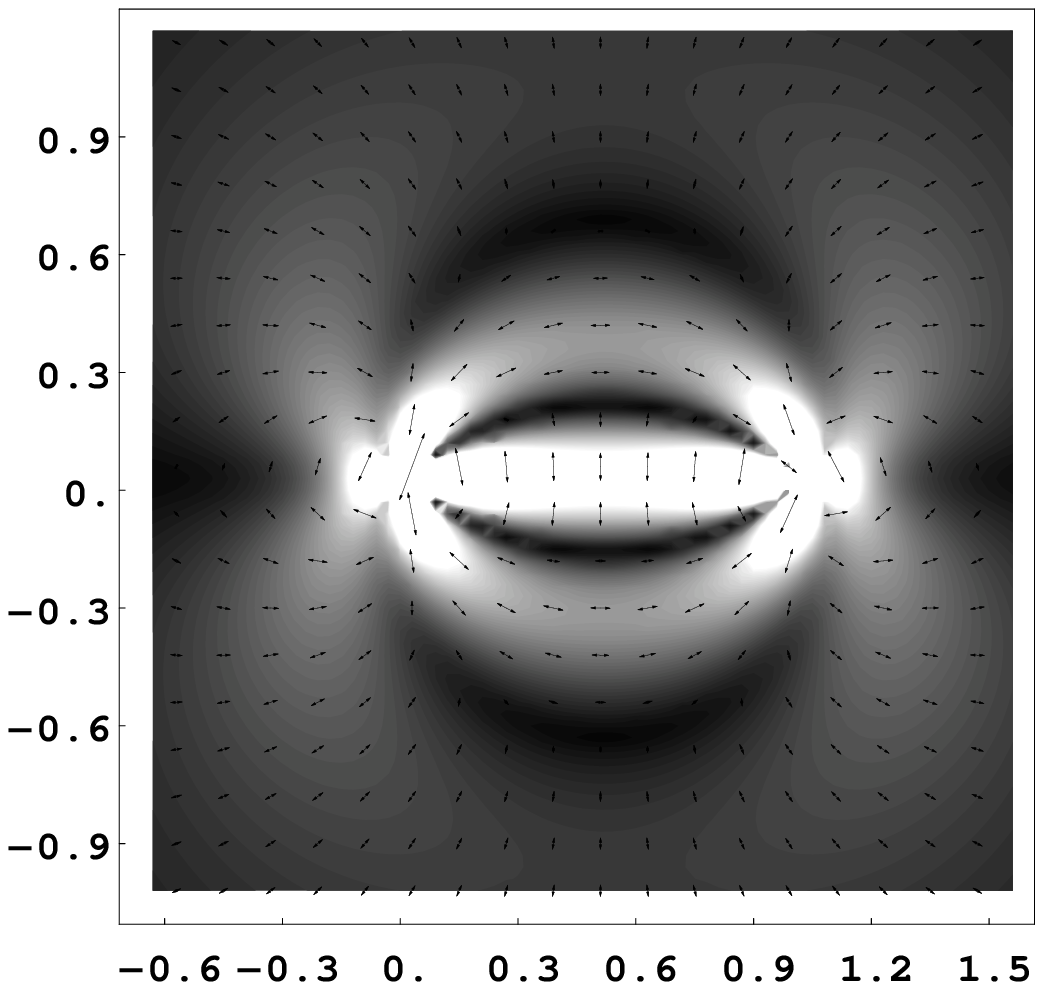}
}
\vspace{.2in}
}
\caption{Pattern of the $\xi_{3}$ field in case of the one-halo model for $l_{3}=1$, $\alpha=-1.$ (left panel) and $\alpha=-0.5$ (right panel). In this plot, $\vr_{1}=(0,0)$ and $\vr_{2}=(1,0)$. They level indicates the amplitude of this component of the shear correlation.}
\label{xi3pattern}\end{figure}

The $\xi_{3}$ patterns are depicted on Fig. \ref{xi3pattern}. They are very similar to what has been observed in numerical simulations. This part of the components does not however encompass the signal coming from equilateral configurations. The extension of the uniform and strong pattern in between the two points depends again on the halo profile. 

\section{Prospects}

The use of complex coordinates proved fruitful in computations of shear correlation functions. It has been possible to obtain the shear three-point correlation functions in manageable forms for the one-halo model with a power law profile.
In the context of this model those results can be used to infer general properties for the shear correlation patterns, as in Benabed \& Scoccimarro (2005); a first step toward the construction of robust methods for the measurement of the shear three-point function. 

The theoretical framework presented in this paper is actually very general and can be used for alternative description of the density correlation properties as such those used in the quasilinear or the nonlinear regime. Indeed the mathematical relation between the projected potential correlation functions and those of the projected density is rather simple and could be integrated out for simple cases\footnote{It is to be noted however that the potential correlations are not uniquely defined from those of the projected density field.}
opening the way for explicit computations of the shear correlation functions in more diverse cases.
 \vskip2pc
\begin{acknowledgements}
\end{acknowledgements}
The calculations presented in this paper would not have been possible without the formula initially excavated by Rom\'an Scoccimarro.


\end{document}